\documentclass{aa}
\usepackage{graphicx}
\usepackage{longtable}
\usepackage{latexsym}
\usepackage{amssymb}
\usepackage{lscape}
\begin{document}
   \title{Optical spectroscopy and the UV luminosity function of galaxies in the 
   Abell 1367, Coma and Virgo clusters.\thanks{Based on observations obtained with the Loiano telescope belonging to the University of Bologna (Italy), 
	with the G.Haro telescope of the INAOE (Mexico) and with the Calar Alto observatory operated by the Centro Astronomico
	Hispano Aleman (Spain).}}

   \author{L.Cortese\inst{1}, G.Gavazzi\inst{1}, J.Iglesias-Paramo\inst{2}, A.Boselli\inst{2} \and L.Carrasco\inst{3,4}}

   \offprints{G.Gavazzi}

   \institute{Universit\'{a} degli Studi di Milano-Bicocca, P.zza della Scienza 3, 20126 Milano, Italy.\\
              \email{Luca.Cortese@mib.infn.it; Giuseppe.Gavazzi@mib.infn.it}
         \and
             Laboratoire d'Astrophysique de Marseille, BP8, Traverse du Siphon, F-13376 Marseille, France.\\
             \email{jorge.iglesias@oamp.fr; alessandro.boselli@oamp.fr}
          \and
	     Instituto Nactional de Astrofisica, Optica y Electronica, Apartado Postal 51 C.P. 72000 Puebla, Pue., Mexico.
	     \email{carrasco@transum.inaoep.mx}
	  \and
	     Observatorio Astronomico Nacional/UNAM, Ensenada B.C., Mexico.	           
             }

   \date{Received 12 November 2002; accepted 14 January 2003}

\abstract{Optical spectroscopy of 93 galaxies, 60 projected in the direction of Abell 1367, 
21 onto the Coma cluster and 12 on Virgo, is reported. The targets were selected either because
they were detected in previous H$\alpha$, UV or $r'$ surveys.  
The present observations bring to 100\% the redshift completeness of H$\alpha$ selected galaxies 
in the Coma region and to 75\% in Abell 1367.
All observed galaxies except one show H$\alpha$ emission and belong to the clusters. 
This confirms previous determinations of the H$\alpha$ luminosity function of the two clusters
that were based on the assumption that all H$\alpha$ detected galaxies were cluster members.
Using the newly obtained data we re-determine the UV luminosity function of Coma and we 
compute for the first time the UV luminosity function of A1367. 
Their faint end slopes remain uncertain (-2.00$<$$\rm \alpha$$<$-1.35)
due to insufficient knowledge of the background counts. 
If 90\% of the UV selected galaxies without
redshift will be found in the background (as our survey indicates), 
the slope of UV luminosity function will be $\alpha$$\sim$-1.35, in
agreement with the UV luminosity function of the field (Sullivan et al. 2000) and with the H$\alpha$ luminosity 
functions of the two clusters (Iglesias-Paramo et al. 2002). 
We discover a  point like H$\alpha$ source in the Virgo cluster, associated with the giant galaxy VCC873,
possibly an extragalactic HII region similar to the one recently observed in Virgo by Gerhard et al. (2002). 
}

\titlerunning{Optical spectroscopy of galaxies in nearby clusters}
\authorrunning{L.Cortese et al.}

   \maketitle
%

\section{Introduction}

Multi-wavelength determinations of the luminosity function of galaxies belonging to nearby
rich clusters are fundamental tools for shedding light on the process of galaxy evolution and, in conjunction
with similar determinations of "field" galaxies, should help unveiling the
role of the environment on galaxy evolution.\\ 
These determinations suffer from insufficient redshift coverage, a crucial parameter for
discriminating the (minority of) cluster members from the background interlopers. \\
Luminosity functions based on narrow band (i.e. H$\alpha$) surveys are less affected by the redshift incompleteness: 
only emission-line galaxies in a narrow recessional velocity range are detected by the limited band--width of the adopted filters.
Complementary redshift information is nevertheless highly recommended in order to reject
background objects whose H$\beta$ or [OIII] lines would fall into the band designed for
containing the redshifted H$\alpha$ line.
Iglesias-Paramo et al. (2002) carried out a deep H$\alpha$ imaging survey of 1 $\times$ 1 degree area 
in Coma and A1367 clusters 
and obtained the first H$\alpha$ luminosity function of nearby clusters of galaxies. 
By that time approximately fifty percent of H$\alpha$ emitting galaxies had no estimate of the recessional velocity. 
With the aim of measuring the remaining redshifts and for confirming the H$\alpha$ emission of these galaxies
we carried out the spectroscopic survey presented in this paper. 
As a side product we also obtained redshifts of several 
UV selected galaxies in Coma and A1367 regions which enable us to re-compute the
UV luminosity functions of the two clusters.\\
The present paper is arranged as follows: Section 2 contains a description of the galaxy selection criteria. 
The observations and the data-reduction procedures are described in Section 3.
The new redshift are presented in Section 4 along with the data on one interesting object in the Virgo cluster. 
The H$\alpha$ and UV luminosity functions of A1367
and Coma are discussed in Section 5. Conclusions are briefly summarized in Section 6.

\section{The sample selection criteria}

\begin{table*}  
\caption{The spectrograph characteristics} 
\label{Tab1} \[ 
\begin{array}{lccccc} 
\hline 
\noalign{\smallskip}   
Observatory & Spectrograph & Dispersion &  Coverage  & CCD & pix \\      
	&  & \rm \AA/mm & \rm \AA &  & \rm \mu m \\ 
\noalign{\smallskip} 
\hline 
\noalign{\smallskip} 
Loiano     & BFOSC & 198 & 3600-8900  & 1340\times1300~EEV & 20 \\
Cananea    & LFOSC & 228 & 4000-7100  & 576\times384~TH & 23 \\
Calar Alto & CAFOS & 187 & 3600-10200 & 2048\times2048~SIT & 24 \\
\noalign{\smallskip} 
\hline 
\end{array} \] 
\end{table*}

Galaxies in the present study were primarily selected among objects, projected in the direction of the Abell 
1367 and Coma clusters, 
with H$\alpha$ emission detected in the INT H$\alpha$ survey of
(\cite{lha})  or selected from the $r'$ survey by \cite{luminosity functionr}.
The H$\alpha$ selected galaxies in the Coma cluster were all spectroscopically measured, while only 75\% of them
were observed in A1367. The $r'$ selected galaxies with $r'\leq17$ have been observed with a completeness
of 64\% and 92\% for A1367 and Coma respectively.\\
In addition to these, we selected 20 targets with UV  
magnitude $\rm m_{uv}<18.5$ 
detected by the FOCA balloon-borne wide field UV camera ($\rm \lambda=2000\AA; \Delta\lambda=150\AA$)  
(Donas et al. 1995 and Donas, private communication).
The UV magnitude is defined as: $\rm m_{uv} = -2.5log(F_{2000}) - 21.175$,
where $F_{2000}$ is the flux in units of $\rm erg~cm^{-2}~s^{-1}~\AA^{-1}.$
The spatial resolution of the UV data is 20 arcsec FWHM (Milliard et al. 1992). The astrometric 
accuracy of these data is therefore typically 3-5 arcsec, insufficient for 
unambiguously identifying spectroscopic targets and discriminating between stars and galaxies. 
To overcome this limitation, we cross-correlated the FOCA catalogues of A1367 and Coma with the $r'$ band 
catalogue of galaxies ($r'<21$) by \cite{luminosity functionr}, using a matching radius of 20 arcsec. 
In cases of multiple identifications we select the galaxy closest to the UV position. Including the present
observations, spectra are available for 64\% and 56\% of the
UV selected galaxies with an optical counterpart
in A1367 and Coma respectively.\\
A dozen of faint H$\alpha$ emitting galaxies in the Virgo cluster with no estimate of the recessional 
velocity in the literature were also selected as filler objects.
These were serendipitously found around bright VCC galaxies by visual inspection to the net ($H\alpha+N[II]$) 
frames obtained at the INT by Boselli et al. (\cite{virgoint}).\\

\section{Observations and data reduction}

Long-slit, low dispersion spectra of 93 galaxies were obtained in several observing runs since 2001 using the 
imaging
spectrograph BFOSC attached to the Cassini 1.5m telescope at Loiano (Italy), LFOSC at the 2.1m telescope of 
the Guillermo
Haro Observatory at Cananea (Mexico), and  with CAFOS attached to the 2.2m telescope of the
Calar Alto Observatory (Spain). Table \ref{Tab1} lists the characteristics of the instrumentation in the 
adopted set-up.\\ 
The observations at Loiano (44 galaxies) were performed using a 2.0 or 2.5 arcsec slit, depending on the seeing conditions, 
generally oriented
E-W. The wavelength calibration was secured with exposures of HeAr lamps. 
The on-target exposure time ranged between 15 and 30 min according to the brightness of the targets.\\
The observations at Cananea (40 galaxies) were carried out with a 1.9 arcsec slit, generally oriented N-S. 
The wavelength calibration was secured with exposures of XeNe lamps. 
The on-target exposure time ranged between
20 and 40 min according to the brightness of the targets.\\
The observations at Calar Alto (9 galaxies) were carried out with a 1.5 arcsec slit, generally oriented N-S. 
The wavelength calibration was secured with exposures of CdHe lamps. The on-target exposure time ranged between
15 and 30 min according to the brightness of the targets. In all runs the observations were obtained in nearly
photometric conditions, with thin cirrus. The orientation of the slit was modified from the set-up given above 
when two
adjacent objects could be observed in the same exposure.\\
The data reduction was performed in the IRAF environment. After bias subtraction, when 3 or more frames of the 
same target 
were obtained, these were combined (after spatial alignment) using a median filter to help cosmic rays removal. 
Otherwise 
the cosmic rays were removed using the task COSMICRAYS and/or under visual inspection. 
The lamps wavelength calibration was checked 
against known sky lines. These were found within $\rm \sim 1~\AA$ from their nominal position, providing 
an estimate of the systematic
uncertainty on the derived velocity of $\rm \sim 50~km~s^{-1}$. 
After subtraction of sky background, one-dimensional spectra were 
extracted from the frames. The redshift were obtained using the cross-correlation 
technique of Tonry \& Davis (\cite{tonry}).
This method is based on a "comparison" between the spectrum of a galaxy (or a star) whose redshift is to be 
determined, and a fiducial 
spectral template of a galaxy of appropriate spectral type to contain the wanted absorption/emission lines. 
The basic assumption behind this method is that the spectrum of a galaxy is well approximated by the spectrum 
of its stars,
modified by the effects of the stellar motion inside the galaxy and by the systemic redshift. For this purpose 
high
signal-to-noise spectra were taken of two template galaxies: VCC0066 (emission lines) and NGC221 
(absorption lines), which were
converted to the rest frame $\rm \lambda$. The derived redshift are not transformed to Heliocentric.

\section{Results}

The velocity measurements obtained in this work are listed in Table 2 as follow:\\
Column 1: Galaxy designation. For Virgo cluster we use the VCC (\cite{vcc}) and VPC (Young \& Currie 1998) 
designations when available.\\
Column 2, 3: (J2000) celestial coordinates, measured with few arcsec uncertainty.\\
Column 4: $r'$ band magnitude.\\
Column 5, 6: observed recessional velocity with uncertainty derived in this work. 
The latter quantity includes only statistical
errors. The global uncertainty can be derived by adding in quadrature $\rm 50~km~s^{-1}$ due to the 
uncertainty in the absolute wavelength calibration.\\
Column 7: type of lines (A=absorption; E=Emission)\\
Column 8: telescope (LOI=Loiano; CAN=Cananea, CAL=Calar Alto)\\
Column 9: galaxies selection band (UV, H$\alpha$ or $r'$)

\subsection{The Virgo cluster}

\begin{figure}
\centering
\includegraphics[width=10.0cm]{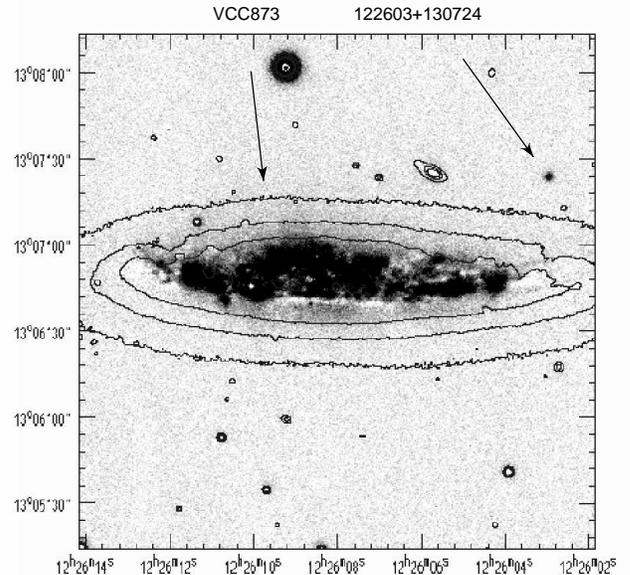}
\caption{Net H$\alpha$+[NII] image of \emph{122603+130724} and VCC873 (grey-scale) with superposed
contours of the continuum emission.}
\label{122603+130724}
\end{figure}

Twelve redshift measurements presented in this work are of faint ($r'\leq18$) objects in the Virgo cluster
which were serendipitously detected  near bright VCC galaxies as part of an H$\alpha$ 
imaging survey of this cluster (Boselli et al. 2002).
Four of these galaxies turned out to be background galaxies whose emission, revealed in 
the net ($H\alpha+N[II]$) frames, is in fact due to H$\beta$ and [OIII] lines, showing that 
at the faint limit of the survey the contamination by this type of objects becomes relevant.\\
We would like to draw the attention on the point-like H$\alpha$ source
\emph{122603+130724} detected at the projected distance of 2' 
from the giant galaxy VCC873 shown in Fig. \ref{122603+130724}.
The source shows an 
H$\alpha$ equivalent width of
$\rm \sim 700\AA$ and an H$\alpha$ flux $\rm \sim 10^{-14.6}~erg~s^{-1}~cm^{-2}$.
In spite of its very faint continuum ($r'\sim21.5$), the 15 min spectrum of this object 
taken with the Loiano 1.5m telescope
yielded a prominent H$\alpha$ emission at  
$\rm 260~km~s^{-1}$, consistent with the velocity of
VCC873 ($\rm 232~km~s^{-1}$).\\
Unlike planetary nebulae (ICPNe) found near giant galaxies in the
Virgo and Fornax cluster (Arnaboldi et al. 1996; Theuns \& Warren 1997) showing strong
[OIII] $\rm \lambda5007\AA$ lines (Dopita et al. 1992),
\emph{122603+130724} is not detected in [OIII].
However, given the weakness of the available spectrum, the upper limit on [OIII] 
is $\sim$ 0.3 of the H$\alpha$ flux, therefore
consistent with \emph{122603+130724} being 
an extragalactic HII region (Dopita et al. 2000) associated with the giant galaxy VCC873.
The  point-like H$\alpha$ source \emph{122544+130740} listed in Table 2
could be another extragalactic HII region, in this case not associated with a bright cluster galaxy. 
However the redshift of this faint ($r'\sim20$) object, based on one line (presumably H$\alpha$)
needs confirmation because it lies at a projected 
distance of 10 arcsec from a bright star.\\
Gerhard et al. (2002) found 17 candidate extragalactic HII regions in the Virgo cluster, among
which the one associated with VCC836 was confirmed spectroscopically. 
Objects of this kind could account for the diffuse intracluster light (Bernstein et al. 1995) 
and might contribute to the enrichment of the intergalactic medium in clusters.

\section{The luminosity functions}
\subsection{The H$\alpha$ luminosity function}
With the new data presented in this work, the redshift of all H$\alpha$ selected galaxies in the Coma cluster 
(\cite{lha}) have been measured. 
All objects turned out to be cluster members ($\rm 4000<V<10000 ~km~s^{-1}$) and significant H$\alpha$ 
emission was detected
in their spectra. The H$\alpha$ luminosity function determined by \cite{lha} assuming that all objects were
at the distance of Coma is fully confirmed.\\
In A1367 we measured the redshift of 14 galaxies, bringing to 75\% the completeness of H$\alpha$ selected 
galaxies in this cluster. 
The membership to the cluster was confirmed for 13 galaxies, while 1 (\emph{114430+195718}) turned out to be 
a background galaxy whose emission, revealed by 
\cite{lha} is in fact due to H$\beta$ and [OIII] lines. 
This confirms that the H$\alpha$ survey by \cite{lha} is little contaminated by 
H$\beta$ and [OIII] emission lines from background objects, emphasizing the high success-rate of 
selecting emission line members with the \cite{lha} method.

\begin{figure}[!t]
\centering
\includegraphics[width=8.5cm]{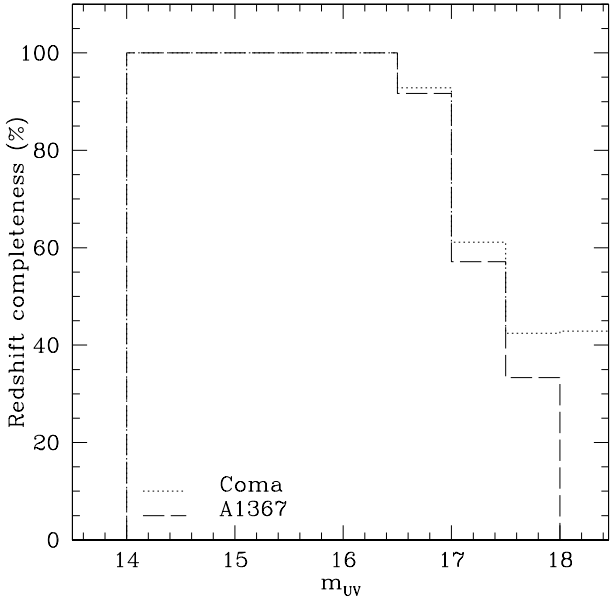}
\caption{Redshift completeness of the UV selected galaxies in Coma (dotted histogram) and A1367 
(long dashed histogram). 
The redshift completeness is $\sim$ 90\% for $m_{UV}\leq$ 16.75 while at fainter magnitudes it 
drops rapidly under 50\%}
\label{Compl}
\end{figure}

\subsection{The UV luminosity function}
\begin{figure}[!t]
\centering
\includegraphics[width=8.5cm]{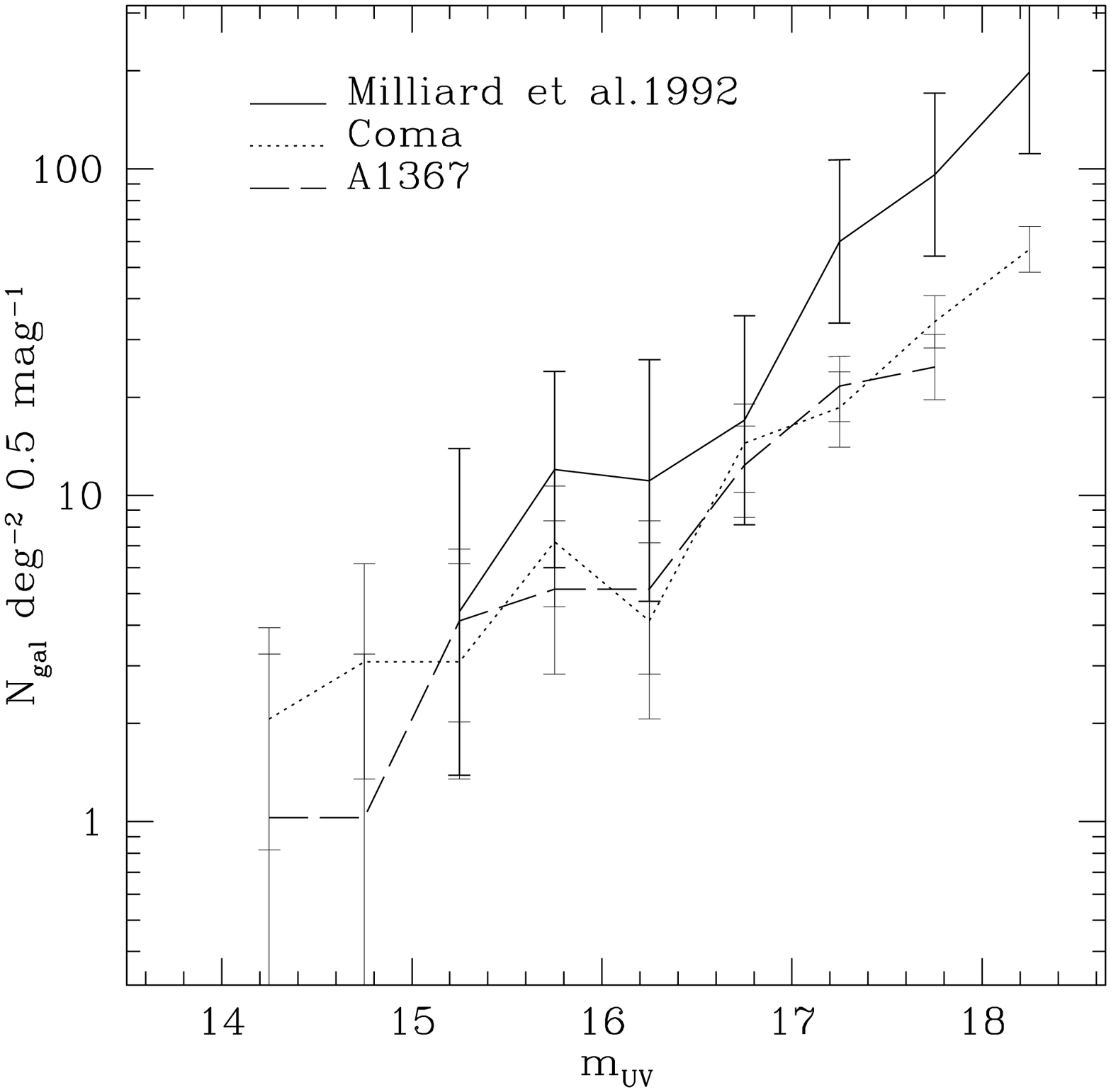}
\caption{Differential UV galaxy counts in the field by Milliard et al. 1992 (solid line), 
in the Coma (dotted line) and A1367 (long dashed line) clusters.}
\label{counts}
\centering
\includegraphics[width=8.5cm]{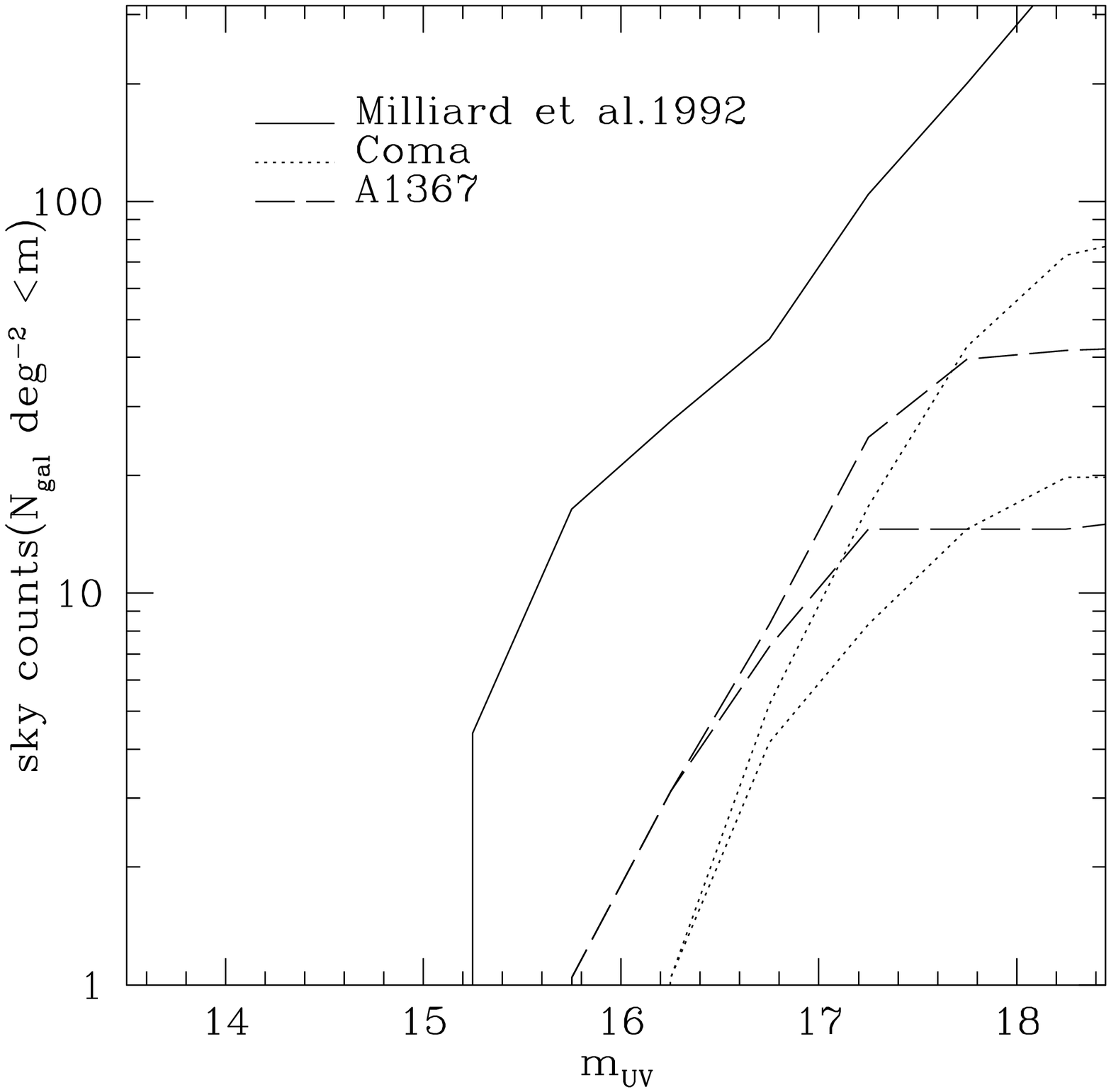}
\caption{Integrated UV galaxy counts in the field by Milliard et al. 1992 (solid line), upper 
and lower background limits for Coma (dotted line) and for A1367 (long dashed line) are given.}
\label{sky}
\end{figure}

\begin{figure}[!t]
\centering
\includegraphics[width=8.5cm]{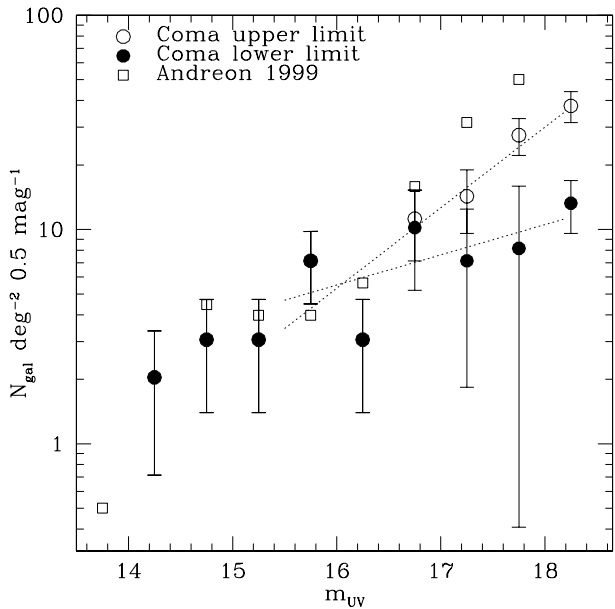}
\caption{The UV luminosity functions of the Coma galaxies. The luminosity function by A99 is represented with 
empty squares and the upper (empty circles) and lower 
(filled points) limits determined in this work are given. The dotted lines represent the best linear fits to
the faint end of the luminosity function. 
Error bars account for Poissonian fluctuations.}
\label{UVcoma}
\end{figure}

The H$\alpha$ luminosity functions of Coma and A1367 are found consistent one another, being characterized by 
a faint end slope 
$\rm \alpha$$\sim$-0.7. This value differs significantly from the slope of the UV luminosity function of Coma 
(Andreon 1999, hereafter A99) 
$\rm \alpha$$\sim$-2.0, the only UV luminosity function ever determined for a cluster. 
While stars with masses M$> 10 M_{\sun}$ and lifetimes $<$ 20 Myr contribute significantly 
to the H$\alpha$ flux, the UV emission is dominated by young stars of intermediate masses 
(2$\rm$ $<$M$<$5 $M_{\sun}$). Furthermore UV emission is
detected also from early-type galaxies with no recent star formation episodes (Deharveng et al. 2002). 
However these differences alone are insufficient for producing discrepant luminosity function such as obtained 
in the two bands.
Thus we use the spectroscopic observations presented in this work to re-compute the Coma UV luminosity function 
and to compute, for the first time, 
the UV luminosity function of A1367. Using the same UV data available to A99, we cross-correlate the UV 
detections with
the $r$' catalogue of \cite{luminosity functionr} for a better star/galaxy discrimination. 
This reduces our luminosity function determination to an area of $\rm 0.97\degr^{2}$ 
(the same over which we determined the H$\alpha$ luminosity function), 
instead of $\rm \sim 3\degr^{2}$ analyzed by A99. 
The determination of the cluster luminosity function requires a reliable estimate of the contribution of 
background/foreground 
objects to the UV counts. This can be accurately achieved for $m_{UV}\leq$ 16.75, since 
for these UV sources the redshift completeness is $\sim$ 90\% (see Fig.\ref{Compl}). 
At fainter magnitudes the redshift completeness drops rapidly, thus requiring the contamination
to be estimated statistically from 
the number of field galaxies, per bin of UV magnitude, that are expected to 
be randomly projected onto the cluster area.
The over-density of UV counts due to Coma/A1367 galaxies can be derived from the difference
between the counts in the direction of the two clusters  
and the UV galaxy counts derived 
in three random fields by \cite{foca} using the same experimental set up as for 
Coma and A1367.
Opposite to our expectation  
the cluster galaxy counts at faint magnitudes are $\sim$ 3 times lower than the \cite{foca} number counts 
(see Fig. \ref{counts}), as already noticed by A99.
The reason for such discrepancy is not fully understood and might reflect a higher degree of star contamination  
among the UV counts than estimated by \cite{foca}. 
\begin{figure}[!t]
\includegraphics[width=8.5cm]{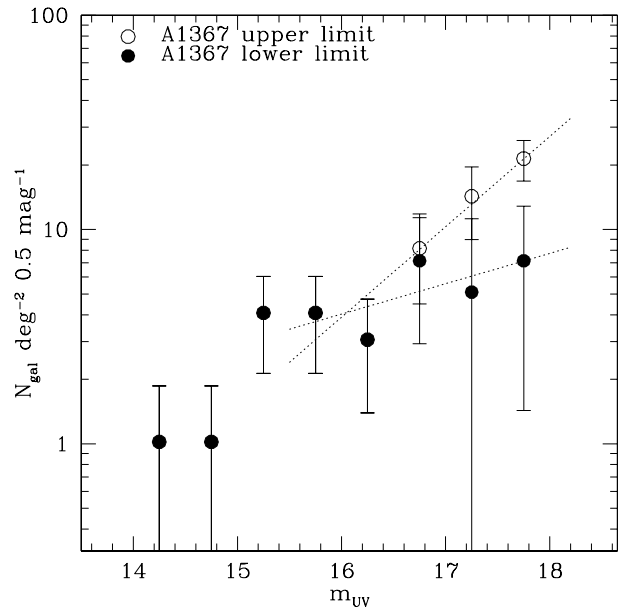}
\caption{Same as Fig. \ref{UVcoma} for A1367.}
\label{UV1367}
\end{figure}
Alternatively it might derive from a higher than expected 
fraction of foreground galaxies in \cite{foca}. However foreground galaxies should contribute
significantly less than background objects due to the small sampled volume at $\rm V<4000~km~s^{-1}$. 
Using our new redshifts we can at least estimate a minimum and a maximum background contribution
to the observed counts in the direction of Coma and A1367, a method followed by A99.
The true background is at least composed of 
all objects whose velocity falls outside the assumed range for the clusters ($\rm 4000<V<10000 ~km~s^{-1}$). 
At most it comprises also all galaxies with unknown redshift.
These limiting estimates, shown in Figure \ref{sky} for 
Coma (dotted line) and A1367 (long dashed line), are found in good reciprocal agreement, but
are significantly lower than the background counts estimated by \cite{foca}.   
The resulting clusters UV luminosity functions should lie in between the two determinations given 
in Figs. \ref{UVcoma} and \ref{UV1367}.\\
Due to the large uncertainties we did not attempt to fit a complete Schechter (1976) function to the data.
Instead we fit the low-end slope (15.5$<m_{UV}<$18) with an exponential form
of slope $km_{UV}$ where $m_{UV}$ is the UV magnitude and $k$ is related to the $\alpha$
parameter of the Schecther function by:
\begin{displaymath}
\alpha = - (k/0.4 + 1)
\end{displaymath}
The resulting UV luminosity function of A1367 shows a slope between $k_{lower}$= 0.14 $\pm$ 0.08 and 
$k_{upper}$ = 0.42 $\pm$ 0.09 (equivalent to $\alpha_{lower}$=-1.35$\pm$0.20 and
$\alpha_{upper}$=-2.05$\pm$0.22). Similarly, for the UV luminosity function of
Coma we find $k_{lower}$ = 0.14 $\pm$ 0.09 and $k_{upper}$ = 0.37$\pm$ 0.09 (equivalent to
$\alpha_{lower}$=-1.35$\pm$0.22 and $\alpha_{upper}$=-1.94$\pm$0.24).
The upper and lower limits of the UV luminosity functions of the two clusters are in fair agreement, 
but the allowed range between
the two limits is still very broad. Likely the low limit gives a more realistic representation of the 
true UV luminosity function because nearly all (18/20 or 
90\%) of the UV selected galaxies for which we obtained a redshift in this work turned out to be 
in the background.
The low limit slope $\alpha$$\sim$-1.35 is in agreement with the field 
UV luminosity function of Sullivan et al. 
(2000) and it is consistent with the slope of the H$\alpha$ luminosity function of the two clusters as determined by
Iglesias-Paramo et al. (2002). 
Certainly the very steep
($\alpha$$\sim$-2.0, -2.2) luminosity function found for Coma by A99 is due to 
an underestimate of the density of background galaxies.

\section{Summary and conclusions}
Optical spectroscopy of 93 galaxies, 60 of which are projected in the direction of Abell 1367, 21 onto the 
Coma cluster and 12 onto Virgo, is reported. These observations bring the redshift 
completeness among H$\alpha$ 
selected galaxies by Iglesias-Paramo et al.(2002) to 100\% in the Coma region and to 75\% 
in Abell 1367. All, except one, the H$\alpha$ selected galaxies show H$\alpha$ emission and are confirmed cluster members.
The exception is a background galaxy whose [OIII] lines fall in the H$\alpha$ filter.
The H$\alpha$ luminosity function of the two clusters determined by 
Iglesias-Paramo et al. (2002) is fully confirmed.\\
Redshifts of UV selected galaxies in Coma and A1367 regions were also obtained. With these new data 
the redshift completeness in the core
of the two clusters has reached $\sim$ 90\% for $m_{UV}\leq$16.75. We re-determine the Coma UV luminosity function 
and we compute for the first time the 
UV luminosity function of A1367. The two are found in fair agreement. Their faint--end slopes remain 
unconstrained (-2.00$<$$\alpha$$<$-1.35) due to the uncertainty on the background counts. However if
90\% of the UV selected galaxies without a
redshift measurement will be found in the cluster background, as our data indicate, the slope of the UV luminosity 
function should be near $\alpha$$\sim$-1.35, in
agreement with the field one (Sullivan et al. 2000) and with the H$\alpha$ luminosity functions of 
the two clusters (Iglesias-Paramo et al. 2002).\\ 
Finally we discover a  point like H$\alpha$ source in the Virgo cluster, 
associated with the giant galaxy VCC873, possibly an extragalactic HII region as the one recently observed 
in Virgo by Gerhard et al.(2002).
Objects of this kind could account for the diffuse intracluster light and might contribute 
to the enrichment of intergalactic medium in galaxy clusters.\\

\begin{acknowledgements}      
We wish to thank Jose Donas for providing us with his unpublished catalogue of UV sources in the Coma and 
Abell 1367 fields. 
The TACS of the Loiano, Cananea and Calar Alto telescopes are acknowledged for the generous time allocation 
to this project. This work could not be completed without the NASA/IPAC Extragalactic Database (NED) which 
is operated by the Jet Propulsion Laboratory, Caltech under contract with NASA.
\end{acknowledgements}

\newpage
\onecolumn
\begin{longtable}{ccccccccc}
\caption{Spectroscopic parameters of the observed galaxies.}\\
\hline 
\noalign{\smallskip}   
     Name &R.A.      &Dec.      & $r'$  & vel          & $ \pm $ & Lines & Tel. & Sel.\\      
          & (J.2000) & (J.2000) & mag & $\rm km \ s^{-1}$ &       &       &      & \\	
\noalign{\smallskip} 
\hline 
\noalign{\smallskip}
\multicolumn{9}{c}{A1367}\\
\hline 
114038+195437 & 11 40 38.96 & +19 54 37.4 & 17.48 &  7784 &  96 & E & LOI & H$\alpha$ \\
114110+201117 & 11 41 10.47 & +20 11 17.7 & 17.57 &  6952 &  22 & E & LOI & H$\alpha$ \\
114117+200832 & 11 41 17.60 & +20 08 32.0 & 15.93 & 14500 &  20 & E & LOI & UV \\
114137+194451 & 11 41 37.90 & +19 44 51.0 & 17.14 & 27891 &  80 & A & CAN & UV \\
114141+200230 & 11 41 41.20 & +20 02 30.5 & 17.37 &  8447 &  51 & E & LOI & H$\alpha$ \\
114142+200054 & 11 41 42.57 & +20 00 54.9 & 17.33 &  8456 &  69 & E & LOI & H$\alpha$ \\
114149+194605 & 11 41 49.79 & +19 46 05.1 & 17.52 &  7789 &  39 & E & LOI & H$\alpha$ \\
114156+194836 & 11 41 57.03 & +19 48 36.2 & 16.53 & 28896 & 140 & A & CAN & $r'$ \\
114229+195238 & 11 42 29.00 & +19 52 38.0 & 17.37 & 14084 & 130 & A & CAN & UV \\
114238+192103 & 11 42 38.40 & +19 21 03.0 & 17.16 & 18674 &  72 & E & LOI & UV \\
114240+195716 & 11 42 40.36 & +19 57 16.6 & 17.68 &  7501 &  46 & E & LOI & H$\alpha$ \\
114252+192543 & 11 42 52.22 & +19 25 43.4 & 15.13 & 24452 & 114 & A & LOI & $r'$ \\
114253+201039 & 11 42 53.11 & +20 10 39.8 & 15.43 &  6261 & 192 & A & CAN & $r'$ \\
114300+201225 & 11 43 00.30 & +20 12 25.9 & 14.90 & 20965 & 133 & A & CAN & $r'$ \\
114306+195620 & 11 43 06.31 & +19 56 20.0 & 15.21 &  6321 & 103 & A & LOI & $r'$ \\
114308+194155 & 11 43 08.00 & +19 41 55.0 & 16.45 & 12988 &  16 & E & LOI & UV \\
114311+200144 & 11 43 11.19 & +20 01 44.9 & 15.37 &  6994 & 102 & A & CAN & $r'$ \\
114312+193841 & 11 43 12.68 & +19 38 41.0 & 16.23 & 27884 & 152 & A & CAN & $r'$ \\
114313+193645 & 11 43 13.08 & +19 36 45.8 & 17.27 &  6121 & 131 & E & LOI & H$\alpha$ \\
114314+194016 & 11 43 14.23 & +19 40 16.7 & 15.03 & 28078 & 142 & A & CAN & $r'$ \\
114315+195614 & 11 43 15.57 & +19 56 15.3 & 16.57 & 28847 & 117 & A & CAN & $r'$ \\
114331+200058 & 11 43 31.30 & +20 00 58.0 & 16.44 & 30164 &  23 & E & LOI & UV \\
114337+191836 & 11 43 37.70 & +19 18 36.0 & 16.65 & 24154 & 132 & E & CAN & UV \\
114341+200135 & 11 43 41.62 & +20 01 35.3 & 17.08 &  6455 &  66 & E & LOI & H$\alpha$ \\
114345+192332 & 11 43 45.41 & +19 23 32.0 & 15.60 & 27889 & 258 & A & LOI & $r'$ \\
114350+192606 & 11 43 50.85 & +19 26 06.2 & 15.31 &  6135 & 141 & A & CAN & $r'$ \\
114353+194321 & 11 43 53.88 & +19 43 21.3 & 17.15 & 24613 & 185 & A & CAN & $r'$ \\
114353+194428 & 11 43 53.77 & +19 44 28.6 & 15.30 &  6120 & 131 & A & CAN & $r'$ \\
114355+192743 & 11 43 55.71 & +19 27 43.9 & 18.72 &  6427 &  50 & E & LOI & H$\alpha$ \\
114357+195607 & 11 43 57.46 & +19 56 07.6 & 15.47 &  7057 &  88 & A & CAN & $r'$ \\
114401+191555 & 11 44 01.10 & +19 15 55.0 & 16.57 & 28353 &  76 & A & CAN & UV \\
114401+191707 & 11 44 01.00 & +19 17 07.0 & 16.92 & 12784 &  91 & E & CAN & UV \\
114401+192953 & 11 44 01.00 & +19 29 53.0 & 16.31 & 19367 &  25 & E & LOI & UV \\
114403+194433 & 11 44 03.31 & +19 44 33.0 & 15.53 &  6715 & 101 & A & CAN & $r'$ \\
114407+193724 & 11 44 07.80 & +19 37 24.0 & 17.28 & 39774 &  85 & A & CAN & UV \\
114413+192012 & 11 44 13.80 & +19 20 12.0 & 16.90 &  5852 &  47 & E & LOI & $r'$ \\
114419+191902 & 11 44 19.30 & +19 19 02.0 & 16.90 & 26091 & 217 & E & CAN & UV \\
114430+195718 & 11 44 30.41 & +19 57 18.8 & 20.23 & 96341 &  19 & E & LOI & H$\alpha$ \\
114444+194814 & 11 44 44.28 & +19 48 14.0 & 19.62 &  8098 &  62 & E & CAL & H$\alpha$ \\
114446+194737 & 11 44 46.13 & +19 47 37.5 & 19.35 &  8240 &  59 & E & CAL & H$\alpha$ \\
114446+194639 & 11 44 46.68 & +19 46 39.5 & 22.03 &  8383 &  36 & E & CAL & H$\alpha$ \\
114447+194648 & 11 44 47.54 & +19 46 48.8 & 21.81 &  8428 &  33 & E & CAL & H$\alpha$ \\
114450+194605 & 11 44 50.81 & +19 46 05.1 & 20.16 &  8089 &  50 & E & LOI & H$\alpha$ \\
114451+194717 & 11 44 51.27 & +19 47 17.5 & 19.24 &  8022 &  35 & E & LOI & H$\alpha$ \\
114454+194733 & 11 44 54.22 & +19 47 33.2 & 18.32 &  8067 &  31 & E & LOI & H$\alpha$ \\
114454+200101 & 11 44 54.55 & +20 01 02.0 & 16.14 &  7646 & 500 & E & CAN & H$\alpha$ \\
114459+194757 & 11 44 59.35 & +19 47 57.1 & 18.42 & 39550 &  79 & E & LOI & $r'$ \\
114501+194550 & 11 45 01.90 & +19 45 50.0 & 16.74 & 20575 &  30 & E & CAN & UV \\
114502+194520 & 11 45 02.71 & +19 45 20.6 & 16.74 & 20626 &  13 & E & LOI & $r'$ \\
114505+194514 & 11 45 05.67 & +19 45 14.8 & 18.07 & 20263 & 150 & A & CAN & $r'$ \\
114506+194733 & 11 45 05.98 & +19 47 33.6 & 16.82 & 19919 &  58 & E & LOI & $r'$ \\
114506+200921 & 11 45 06.56 & +20 09 21.4 & 15.36 &  6145 & 184 & A & LOI & $r'$ \\
114513+194523 & 11 45 13.77 & +19 45 22.2 & 15.84 &  8316 & 224 & E & CAN & H$\alpha$ \\
114514+200836 & 11 45 14.10 & +20 08 36.0 & 16.99 &  3752 &  43 & E & CAN & UV \\
114518+200009 & 11 45 18.00 & +20 00 09.5 & 17.54 &  5327 &  28 & E & LOI & H$\alpha$ \\
114525+194905 & 11 45 25.56 & +19 49 05.7 & 15.22 &  8422 & 123 & A & CAN & $r'$ \\
114541+194613 & 11 45 41.16 & +19 46 13.4 & 14.89 &  5419 & 138 & A & CAN & $r'$ \\
114545+194130 & 11 45 45.20 & +19 41 30.0 & 16.48 &  6123 &  17 & E & CAN & UV \\
114558+194810 & 11 45 58.59 & +19 48 10.9 & 15.41 &  5493 & 173 & A & LOI & $r'$ \\
114611+195110 & 11 46 11.99 & +19 51 10.0 & 15.07 &  5441 & 101 & A & CAN & $r'$ \\
\hline
\multicolumn{9}{c}{Coma}\\
\hline
\noalign{\smallskip}
125802+282722  & 12 58 02.10 & +28 27 22.0 & 17.01 & 48300 & 150 & A & CAN & UV \\
125815+283117  & 12 58 15.12 & +28 31 17.2 & 16.70 &  6725 &  56 & E & LOI & UV \\
125823+281945  & 12 58 23.66 & +28 19 45.6 & 16.26 &  8352 & 126 & A & CAN & $r'$ \\
125828+283135  & 12 58 28.45 & +28 31 35.9 & 15.24 & 18290 & 104 & A & CAN & $r'$ \\
125829+281806  & 12 58 29.44 & +28 18 06.3 & 15.52 &  6082 &  99 & A & CAN & $r'$ \\
125845+283235  & 12 58 45.80 & +28 32 35.3 & 17.76 &  6333 &  14 & E & LOI & H$\alpha$ \\
125845+284133  & 12 58 45.64 & +28 41 33.1 & 17.21 &  6682 & 149 & E & LOI & H$\alpha$ \\
125856+282749  & 12 58 56.19 & +28 27 49.0 & 15.23 &  5927 & 156 & A & CAN & $r'$ \\
125918+283726  & 12 59 17.90 & +28 37 26.0 & 16.81 & 10340 &  32 & E & CAN & UV \\
125923+282918  & 12 59 23.06 & +28 29 18.4 & 15.51 &  7015 & 165 & E & CAN & H$\alpha$ \\
125924+282050  & 12 59 24.11 & +28 20 50.1 & 16.62 & 20241 &  26 & E & LOI & UV \\
125941+283026  & 12 59 41.12 & +28 30 26.6 & 15.45 &  8314 & 161 & A & CAN & $r'$ \\
125956+275548  & 12 59 56.68 & +27 55 48.2 & 15.31 &  7653 &  93 & A & CAN & $r'$ \\
130004+283614  & 13 00 04.51 & +28 36 14.1 & 15.32 &  6727 & 153 & A & CAN & $r'$ \\
130037+283951  & 13 00 37.22 & +28 39 51.7 & 16.62 &  7095 &  87 & E & CAN & H$\alpha$ \\
130127+281053  & 13 01 27.29 & +28 10 53.9 & 17.18 & 48821 & 100 & A & CAN & $r'$ \\
130128+281515  & 13 01 28.63 & +28 15 15.9 & 20.41 &  9061 &  33 & E & LOI & H$\alpha$ \\
130130+283328  & 13 01 30.85 & +28 33 28.0 & 16.76 &  6885 & 194 & E & LOI & H$\alpha$ \\
130140+281456  & 13 01 40.97 & +28 14 56.6 & 19.33 &  9286 &  20 & E & LOI & H$\alpha$ \\
130151+280822  & 13 01 51.90 & +28 08 22.0 & 17.00 & 28754 &  27 & A & CAL & UV \\
130158+282114  & 13 01 58.43 & +28 21 14.8 & 19.81 &  8577 &  18 & E & LOI & H$\alpha$ \\
\hline
\multicolumn{9}{c}{Virgo}\\
\hline
VCC1715        & 12 37 28.45 & +08 47 39.2 & 15.37  &	888 &  32 & E & LOI & H$\alpha$	\\
VCC 404        & 12 20 17.50 & +04 12 10.0 & 14.20  &  1746 &  93 & E & LOI & H$\alpha$	\\
VPC766         & 12 30 46.20 & +12 05 57.1 & 17.05  &  1162 &  47 & E & LOI & H$\alpha$	\\
122544+130740  & 12 25 44.20 & +13 07 40.0 & 20.10  &   796 &  37 & E & CAL & H$\alpha$	\\
122603+130723  & 12 26 02.91 & +13 07 23.7 & 21.50  &	260 &  34 & E & LOI & H$\alpha$	\\
122605+130725  & 12 26 05.75 & +13 07 25.7 & 17.14  & 23980 &  40 & E & LOI & $r'$	\\
123015+122812  & 12 30 15.20 & +12 28 12.0 & 21.10  & 79521 &  37 & E & CAL & H$\alpha$	\\
123021+121614  & 12 30 21.10 & +12 16 14.0 & 20.70  & 90792 &  36 & E & CAL & H$\alpha$	\\
123222+090944  & 12 32 22.81 & +09 09 44.1 & 17.97  & 77812 &  50 & E & LOI & H$\alpha$	\\
123315+091448  & 12 33 15.25 & +09 14 48.0 & 17.59  & 30923 &  29 & E & LOI & $r'$	\\
123728+082540  & 12 37 28.00 & +08 25 40.0 & 20.60  & 79521 &  33 & E & CAL & H$\alpha$	\\
124547+102620  & 12 45 47.32 & +10 26 20.5 & 16.41  & 36946 & 144 & E & LOI & $r'$	\\
\noalign{\smallskip} 
\hline 
\end{longtable}

\end{document}